\documentclass[twocolumn,showpacs,preprintnumbers,amsmath,amssymb]{revtex4}

\usepackage{graphicx}
\usepackage{dcolumn}
\usepackage{bm}

\begin{document}

\preprint{Version published in PHYSICAL REVIEW D {\bf 78}, 101302(R) (2008)}

\title{Prompt high-energy neutrinos from gamma-ray bursts in photospheric and synchrotron self-Compton scenarios}

\author{Kohta Murase}
\affiliation{%
Yukawa Institute for Theoretical Physics, Kyoto University,
Kyoto, 606-8502, Japan
}%

\date{November 19, 2008}
                        
\begin{abstract} 
We investigate neutrino emission from gamma-ray bursts
(GRBs) under alternative scenarios for prompt emission (the
photospheric and synchrotron self-Compton scenarios)
rather than the classical optically thin synchrotron scenario.
In the former scenario, we find that neutrinos from the $pp$ reaction can
be very important at energies $\lesssim (10-100)$ TeV. They may be
detected by IceCube/KM3Net and
useful as a probe of baryon acceleration around/below the 
photosphere. In the latter scenario, 
we may expect $\sim$ EeV $p \gamma$ neutrinos produced by soft photons. 
Predicted spectra are different from that in the classical scenario, and neutrinos 
would be useful as one of the clues to the nature 
of GRBs (the jet composition, emission radius, magnetic field and so on). 
\end{abstract}

\maketitle
\section{\label{sec:level1}Introduction}
Prompt high-energy neutrino emission from gamma-ray bursts (GRBs) was 
predicted in the internal shock (IS) model \cite{Pac1,Wax1} 
and has been studied by many authors \cite{KM1,Der1,Rac1} 
(for afterglows, see Ref. \cite{KM2} and references therein). 
However, despite recent progresses in the \textit{Swift} era, 
the mechanism of prompt emission has not been well understood \cite{Mes1}. 
One of the most frequently discussed scenarios as 
a standard one is the classical optically thin synchrotron (including diffusive
synchrotron) scenario, where prompt photons around the peak energy in
the hard x-ray or gamma-ray band come from electrons accelerated
at internal shocks and/or by magnetic reconnection in the optically
thin region. 
However, this scenario cannot satisfactorily explain some observational
features such as the lower energy spectral index and 
observed spectral correlations \cite{Pre1}, which may also be 
related to the cooling and efficiency problems \cite{Iok1}. 
Motivated by these problems, photospheric emission models have
been developed \cite{Tho1,Ree1}, which have 
an advantage to stabilize the peak energy \cite{Tho1,Pee1}. 

Another alternative scenario is the synchrotron self-Compton (SSC)
scenario, which may provide more viable parameter sets for prompt emission, 
compared to the classical scenario \cite{Mes2}. 
The synchrotron peak is in the optical/ultraviolet range and gamma-ray 
photons arise from inverse Compton scatterings.  
This scenario may explain some bursts such as GRB 080319B \cite{Kum1},
despite the lack of bright optical emission in many bursts \cite{Rom1}. 
      
In this work, we investigate neutrino emission under
photospheric and SSC scenarios. Our method of calculation based
on Geant4 is basically the same as in
Refs. \cite{KM1,KM2}, but improved qualitatively and
quantitatively by including the $pp$ reaction and 
all the relevant processes of protons, mesons and
muons \cite{Cho1,Mes3}. As for GRB neutrinos, 
cooling of mesons and muons is remarkably important, which 
makes neutrino spectra complicated and affects estimate of event rates 
\cite{Wax1,Rac1,Mes3}. Especially, we demonstrate that 
\textit{pp} neutrinos can be 
important below $\sim 100$ TeV.

\section{\label{sec:level2}The Photospheric Scenario}
GRB prompt emission is considered to be radiation 
from a highly relativistic jet
toward us, and the typical Lorentz factor of 
the collimated outflow is $\Gamma \sim {10}^{2-3}$. 
In the standard picture, a significant fraction of the outflow 
energy is converted to the radiation energy 
via some dissipation mechanism within the outflow 
(e.g., internal shocks or magnetic reconnection), leading 
to the observed isotropic photon luminosity $L_{\gamma} \sim 
{10}^{52}~{\rm ergs}~{\rm s}^{-1}$ \cite{Mes1}.  
In photospheric emission models, the prompt emission comes
from around the photospheric radius $r \sim r_{\rm{ph}}$, 
at which the Thomson optical depth $\tau_{\gamma e} 
\simeq n_e \sigma_{T} (r/\Gamma)$ is unity.  
The photospheric radius is given by $r_{\rm{ph}} \approx (L_{M} \sigma_T/4 \pi \Gamma^3
m_p c^3) \simeq 1.2 \times {10}^{12} \, {\rm{cm}} \, L_{M,52.5}
\Gamma_{2.5}^{-3}$, when the outflow is baryonic ($n_e \approx n_p$).
Here $L_M$ is the isotropic outflow luminosity carried by baryons, 
in the observer frame \cite{Mes1,Tho1}.  
Possibly, the outflow may be dominated by pairs. If $n_e/n_p \sim
m_p/m_e$, we obtain $r_{\rm{ph}} \approx (L_{M} \sigma_T/4 \pi \Gamma^3
m_e c^3) \simeq 2.2 \times {10}^{15} \, {\rm{cm}} \, L_{M,52.5}
\Gamma_{2.5}^{-3}$ \cite{Ree1}. In this scenario, 
dissipation and thermalization should occur around/below the photosphere.
For instance, in the dissipative photospheric model \cite{Ree1}, we 
can assume that the dissipation is maintained after the coasting radius, 
leading to the temperature $k T_{\rm{ob}} \sim 100$ keV in the
observer frame. The observed peak energy $\varepsilon_{\rm{ob}}^b \sim 500$
keV can be achieved by the Comptonized thermal photons \cite{Pee1}.

As an example, let us consider an internal collision of two subshells 
in the relativistic outflow, with $\Gamma_s \sim {10}^{2}$ and $\Gamma_f \sim
{10}^{3}$. The Lorentz factor of the merged subshell is estimated 
as $\Gamma \approx \sqrt{\Gamma_f {\Gamma}_s} \simeq 
{10}^{2.5}$, and the Lorentz factor of the internal shock 
in their center of mass frame is $\sim \sqrt{\Gamma_{\rm{sh}}} \approx
\sqrt{(\Gamma_f/\Gamma_s+\Gamma_s/\Gamma_f)/2} \sim$ a few. 
Hence, internal shocks would be mildly relativistic shocks, at which 
electrons can be accelerated, and prompt emission 
mainly comes from the shocked subshells which could be 
magnetized via plasma instabilities at collisions \cite{Mes1}.  
Not only electrons but also baryons can be 
accelerated even around/below the photosphere \cite{Iok1,Mes3}.
The acceleration time scale of protons in the comoving frame 
of the merged subshells is written as 
$t_{\rm{acc}} = \eta \varepsilon_{p}/eBc$. Although  
realistic values of $\eta$ are not well known, we could expect $\eta \sim (1-10)$ 
in the most efficient cases, 
when we consider the Fermi acceleration mechanism in the Bohm limit 
\cite{Rac1,Gal1}. Here we assume $\eta \sim 10$ as in Refs. \cite{KM1,Rac1}. 
The magnetic field is written as $B \approx 8.4 \times {10}^{5}
\, {\rm{G}} \, \xi_B^{1/2} L_{\gamma, 52}^{1/2} \Gamma_{2.5}^{-1}
r_{\rm{ph},12.5}^{-1}$, where $\xi_{B} \equiv U_B/U_{\gamma}$. 
Although $\xi_B$ is uncertain, we expect that
the equipartition between the magnetic energy density
$U_B$ and photon energy density $U_{\gamma}$ can be achieved in the photospheric 
scenario, as often expected in the classical one. 
The maximum energy of protons is set by the condition 
$t_{\rm{acc}}< {\rm{min}}[t_{p},t_{\rm{dyn}}]$. Here $t_{\rm{dyn}}
\approx r/\Gamma c$ is the dynamical time scale and $t_{p}^{-1} \equiv 
t_{\rm{BH}}^{-1} + t_{p\gamma}^{-1} +  t_{p p}^{-1} + t_{\rm{syn}}^{-1} 
+ t_{\rm{IC}}^{-1} + t_{\rm{ad}}^{-1} + t_{\rm{esc}}^{-1}$ is the
proton loss time scale (in the comoving frame), where $t_{\rm{BH}}$, $t_{p\gamma}$, 
$t_{p p}$, $t_{\rm{syn}}$, 
$t_{\rm{IC}}$, $t_{\rm{ad}}$, and $t_{\rm{esc}}$ are cooling times of the
Bethe-Heitler process, photomeson production, 
\textit{pp} reaction, synchrotron emission, inverse Compton emission, 
adiabatic expansion, and the escape time in the Bohm diffusion 
approximation \cite{Rac1,KM2}. Following Refs. \cite{KM1,KM2}, 
we evaluate the maximum proton energy for each parameter set, whose typical
values are $E_p^{\rm{max}} \sim {10}^{8-10}$ GeV, in the observer frame.

High-energy protons can interact with photons and protons, producing
mesons and muons that decay to neutrinos. 
As in Refs. \cite{Wax1,KM1,Der1,KM2}, it is often convenient to use 
the effective photomeson production optical depth, $f_{p\gamma} \approx 
t_{\rm{dyn}}/t_{p\gamma}$. By using the $\Delta$-resonance approximation at 
$E_p^b \sim 50 \, {\rm{PeV}} \, \Gamma_{2.5}^2 / 
\varepsilon_{\rm{ob}, 316 \, \rm{keV}}^b$ \cite{Wax1,KM1,Der1,KM2}, we have 
$f_{p\gamma}(E_p^b) \simeq 23 \frac{L_{\gamma,51.5}^b \Gamma_{2.5}}{L_{M,52.5}
\varepsilon_{\rm{ob}, 316 \, \rm{keV}}^{b}} \tau_{\gamma e}$, 
where $L_{\gamma}^{b} (< L_{\gamma}$) is the observed 
photon luminosity at $\varepsilon_{\rm{ob}}^b$.
It implies that almost all the energy of high-energy protons can be used 
for photomeson production when $t_{p \gamma}$ is the most important,
and accelerated protons will be depleted.
Similarly, the effective optical depth for \textit{pp} reaction is
$f_{pp} \approx \kappa_{pp} n_p \sigma_{pp} l \simeq 0.05
\tau_{\gamma e}$, where $\sigma_{pp} \simeq 5 \times {10}^{-26} \, \rm{cm}^{-2}$ and 
$\kappa_{pp} \simeq 0.5-0.6$. 
The meson production efficiency $f_{\rm{meson}}$ can
be approximated by $\sim {\rm{min}}[1,{\rm{max}}(f_{p\gamma},f_{pp})]$, 
as long as $t_{p \gamma}$ or $t_{pp}$ is the most relevant loss time
scale. We can expect that a significant fraction of proton energy is used for 
meson production, which is demonstrated through numerical calculations.

\section{\label{sec:level3}The SSC Scenario}
SSC emission models have often been discussed \cite{Mes2}, where 
the observed peak energy $\varepsilon_{\rm{ob}}^b \sim 500$ keV is
identified with the second peak formed by up-scattered synchrotron
photons. Rather large emission radii close to the deceleration radius,
$r_{\rm{dec}} \sim {10}^{16} \, {\rm{cm}} \, (\gg r_{\rm{ph}} \sim
{10}^{12}$ cm) are expected \cite{Mes2,Kum1}. 
The first synchrotron peak in the observer frame is estimated as 
$\varepsilon_{\rm{ob}}^{b1} \approx 
\hbar \gamma_{e,m}^2 (\Gamma eB/m_e c)
\sim 20 \, \, {\rm{eV}} \epsilon _{e,-1}^2 \xi_{B,-2}^{1/2}  
L _{\gamma,52}^{1/2} r_{15.5}^{-1}$,
and then the second peak coming from inverse Compton scatterings is 
$\varepsilon_{\rm{ob}}^{b2} \approx \gamma_{e,m}^2 \varepsilon_{\rm{ob}}^{b1}
\sim 700 \, \, {\rm{keV}} \epsilon _{e,-1}^4 \xi_{B,-2}^{1/2} 
L_{\gamma,52}^{1/2} r_{15.5}^{-1}$. In addition, the third peak of
$\varepsilon_{\rm{ob}}^{b3} \sim 20$ GeV can also be expected, 
which is one of the predictions in this scenario. Most of the
radiation energy can be released as high-energy gamma rays, so that 
we expect $\mathcal{E}_{\gamma}^{\rm{iso}} \simeq Y 
\mathcal{E}_{\gamma 2}^{\rm{iso}}$ with $Y \sim \xi_{B}^{-1/2}$, where 
$Y$ is the Compton Y parameter \cite{Mes1} and 
$\mathcal{E}_{\gamma 2}^{\rm{iso}}$ is the isotropic energy around the 
second peak. 

Note that, in this scenario, the magnetic field should be smaller than in
the equipartition case. Corresponding to $Y \sim 10$, 
$\xi_{B} \equiv U_B/U_{\gamma} \sim {10}^{-2}$ leads to $B \sim 46 \,
{\rm{G}} \, \xi_{B,-2}^{1/2} L_{\gamma, 52}^{1/2} \Gamma_{2.5}^{-1}
r_{15.5}^{-1}$. Then, from the condition 
$t_{\rm{acc}} < {\rm{min}} [t_{p}, t_{\rm{dyn}}]$, the 
maximum proton energy is typically 
estimated as $E_{p}^{\rm{max}} \sim {10}^{10-11}$ GeV for $\eta \sim 10$. 
Production of ultra-high-energy cosmic rays (UHECRs) may be possible. 

We can also evaluate the effective photomeson optical depth as
$f_{p\gamma}(E_p) \simeq 9 \frac{L_{\gamma,50.5}^{b1}}{r_{15.5} \Gamma _{2.5}^2
\varepsilon_{\rm{ob}, 32 \, \rm{eV}}^{b1}}
{(E_p/E_p^{b1})}^{\beta-1}$, 
at sufficiently high energies below $E_{p}^{b1} \sim 500 \, {\rm{EeV}}
\, \Gamma_{2.5}^2/ \varepsilon_{{\rm{ob}}, 32 \, {\rm{eV}}}^{b1}$.
When $t_{p \gamma}$ is the most relevant, we expect a 
significant fraction of proton energy is used
for the photomeson production around the highest energies. 
But, in this scenario, $f_{p \gamma}$ and $f_{pp}$ are rather small at lower
energies, so that 
efficient $\lesssim$ PeV neutrino production is not expected. 

\section{\label{sec:level4}The Neutrino Spectrum and Flux}
In this work, we numerically calculate neutrino spectra and evaluate
fluxes through the method used in Refs. \cite{KM1,KM2}. 
Calculations are first executed in the comoving frame, and then results in the observer 
frame can be easily obtained via the transformation. 
The photomeson production and \textit{pp} reaction are treated in detail,  
as well as various cooling processes of mesons and muons, 
i.e., synchrotron, inverse Compton, meson-photon, 
$\pi p/ \mu p$ and adiabatic cooling processes.
Neutrino spectra can be calculated, giving a target photon spectrum with
$U_{\gamma}=(L_{\gamma}/4 \pi r^2 \Gamma^2 c)$, target (thermal)
proton density $n_p = (U_{\rm{th}}/m_p c^2)$ and $U_{B} = \xi_{B}
U_{\gamma}$. For the evaluation of neutrino fluxes, proton fluxes should also 
be given and we use $dn_{\rm CR}/d \varepsilon _{p} \propto 
\varepsilon _{p}^{-2}$ in this work. Although the proton spectral index of
$p \sim 2$ is often expected for nonrelativistic or mildly 
relativistic shocks with the compression ratio of $\sim 4$, 
different values are possible, for example, due to large angle scatterings rather 
than small pitch-angle scatterings across relativistic shocks \cite{Gal1,Kir1}. 
The baryon energy input $\mathcal{E}_{\rm{CR}}$ is 
given by using the nonthermal baryon loading factor 
$\xi_{\rm{acc}} \equiv \mathcal{E}_{\rm{CR}} /
\mathcal{E}_{\gamma} =  \mathcal{E}_{\rm{CR}}^{\rm{iso}} /
\mathcal{E}_{\gamma}^{\rm{iso}}$, where
$\mathcal{E}_{\gamma}$ and $\mathcal{E}_{\gamma}^{\rm{iso}}$ are the
geometrically corrected radiation energy and isotropic radiation
energy, respectively. Here we adopt
$\xi_{\rm{acc}} \sim 1-10$, as in Refs. \cite{Wax1,KM1,Der1,KM2}.
Differences in spectra mainly come from 
$r$, $\xi_{B}$ and the photon spectrum, which depend on prompt 
emission scenarios. For instance, in the classical scenario, 
$r \sim {10}^{13-15}$ cm, $\xi_{B} \sim 1$ and the broken 
power-law photon spectrum are typically used \cite{Wax1,KM1,Der1,Rac1}. 

In the photospheric scenario, we may expect smaller radii of $r \sim
{10}^{11-13}$ cm and strong magnetic fields with $\xi_{B} \sim 1$.
Prompt emission comes from the photosphere where $\tau_{\gamma
e}=1$, so that we can adopt the broken power-law photon spectrum 
as $dn/d\varepsilon  \propto {(\varepsilon/
{\varepsilon}^{b})}^{-\alpha}$ for $\varepsilon^{\rm{min}} 
< \varepsilon < \varepsilon ^{b}$ and $dn/d\varepsilon \propto
{(\varepsilon/{\varepsilon}^{b})}^{-\beta}$ for $\varepsilon
^b < \varepsilon < \varepsilon ^{\rm{max}}$, expressed 
in the comoving frame.
Here we set $\varepsilon ^{\rm{min}}=1$ eV 
and $\varepsilon ^{\rm{max}}=1$ MeV,  
because the synchrotron self-absorption and 
pair-creation absorption will be crucial below and 
above these energies, respectively.
But our results are not sensitive to them.
We take $\varepsilon^{b}=\varepsilon_{\rm ob}^{b}/\Gamma = 1$ keV, 
$\alpha=1$ and $\beta=2.2$, which are obtained from observations.  
However, since photons would be significantly thermalized when 
$\tau_{\gamma e} > 1$, we adopt the black-body
photon spectrum with the temperature $T ={(U_{\gamma}/a)}^{1/4}$ for
subphotospheric emission. 
$n_p$ is given assuming $U_{\rm{th}}=U_{\gamma}$,
and we use $\tau_{\gamma e}$ as a parameter instead of $L_{\gamma}^b$. 

\begin{figure}[t]
\includegraphics[width=\linewidth]{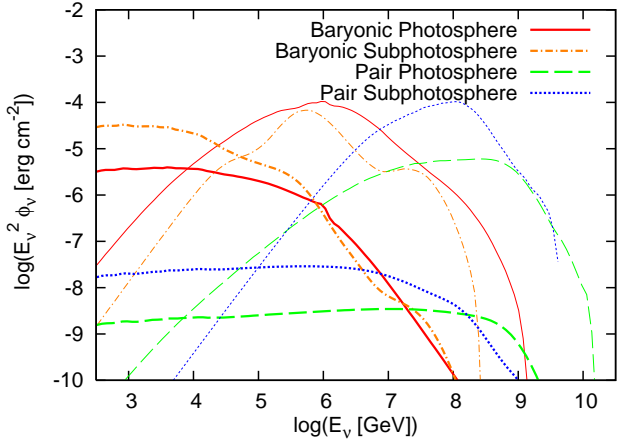}
\caption{\small{\label{Fig1} 
The $(\nu _{\mu} + \bar{\nu} _{\mu})$ fluence from a GRB event at $z=0.1$. 
\textit{pp} neutrinos (thick line) and $p\gamma$ neutrinos (thin line) are shown.
Baryonic photosphere: $r=r_{\rm{ph}}={10}^{12.5}$ cm 
($\tau_{\gamma e}=1$); $N_{\mu} \sim 1.7$ events. Baryonic subphotosphere:
$r={10}^{-0.5} r_{\rm{ph}}$ ($\tau_{\gamma e}=10$); $N_{\mu} \sim 1.4$
events. Pair photosphere: $r=r_{\rm{ph}}={10}^{14.2}$ cm
($\tau_{\gamma e}=1$); $N_{\mu} \sim 0.018$ events. Pair subphotosphere:
$r={10}^{-0.5} r_{\rm{ph}}$ ($\tau_{\gamma e}=10$); $N_{\mu} \sim
0.16$ events. $\mathcal{E}_{\gamma}^{\rm{iso}} = {10}^{53.5} \, \rm{ergs}$, 
$\Gamma={10}^{2.5}$, $\xi _{B}=1$, and $\xi_{\rm{acc}}=1$ are adopted. 
We assume $n_e=n_p$ in case of the baryonic photosphere
while $n_e = \frac{m_p}{m_e} n_p$ in possible cases of the pair photosphere.
}}
\end{figure}

In the SSC scenario, we expect larger radii of $r \sim
{10}^{15-16}$ cm and relatively weak magnetic fields with $\xi_{B} \sim 0.01$.
The photon spectrum is given by the sum of multi-broken power-law 
spectra with break energies,
$\varepsilon^{b1} = ({10}^{-2}-{10}^{-1})$ eV and $\varepsilon^{b2} = 1$
keV with $\alpha=1$ and $\beta =2.2$. 
Note that larger emission radii lead to lower synchrotron self-absorption energies 
of $\varepsilon^{\rm{min}} = ({10}^{-3}-{10}^{-1})$ eV.
For comparison, we also show neutrinos produced via
the $p \gamma$ reaction between protons accelerated at the external
reverse-shock (RS) and prompt photons produced by internal dissipation.
The relevant quantities at the crossing time, such as the radius $r_{\times}$, 
Lorenz factor of the shocked ejecta $\Gamma_{\times}$ 
and magnetic field $B_{\times}$, are evaluated according to 
the standard reverse-forward shock theory \cite{Mes1}. The 
procedure is described in Ref. \cite{KM2}. 

The results in the photospheric scenario are shown in Fig. 1.  
We can see a \textit{pp} neutrino component is dominant
at $E_{\nu} \lesssim 10$ TeV, and more important for 
subphotospheric emission.  
The Bethe-Heitler and photomeson production processes become 
more relevant at higher energies. The former can lead to 
a dip between $pp$ and $p \gamma$ components 
(dotted-dashed lines). The latter makes
$p\gamma$ neutrinos, whose fluence is suppressed at sufficiently high
energies since mesons and muons cool before they decay. 
For example, we can find break energies around PeV, and 
$p \gamma$ neutrinos from kaons at higher energies $E_{\nu} \gtrsim
10$ PeV, for baryonic photosphere and baryonic subphotosphere.  
When the subphotospheric emission is expected, dissipation 
would continue from the subphotospheres all the way to the photosphere,
enabling us to expect that \textit{pp} neutrinos from subphotospheres are more 
important at lower energies, while $p \gamma$ neutrinos from 
around/above the photosphere at higher energies. 
Muon event rates expected by IceCube, $N_{\mu} (>{10}^{2.5} \,
\rm{GeV})$, are also shown in the figure caption. If
a burst occurs at $z \lesssim 0.1$, we may detect a few events. 
Event rates of $pp$ neutrinos dominate over those of $p \gamma$ 
neutrinos for baryonic subphotosphere. 

\begin{figure}[t]
\includegraphics[width=\linewidth]{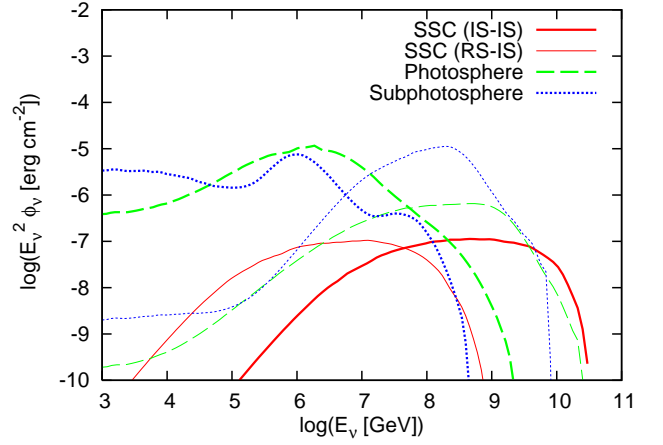}
\caption{\small{\label{Fig2}
As in Fig. 1, but $z=0.9$.
SSC (IS-IS) (thick solid line): $r ={10}^{16}$ cm, $\Gamma={10}^{3}$, 
$L_{\gamma}^b ={10}^{52.3}\, \rm{ergs} \, {s}^{-1}$, 
and $\xi_B=0.01$; $N_{\mu} \sim 2 \times {10}^{-4}$ events. 
SSC (RS-IS) (thin solid line): $r_{\times} \simeq {10}^{16.4}$ cm, 
$\Gamma_{\times} \simeq 150$, and $B_{\times}^{r} \simeq 7.7$ G
(coming from $E_{\rm{ej}}^{\rm{iso}}=4 \times {10}^{54}$ ergs,
$\Delta_0=7.5 \times {10}^{11}$ cm, $\epsilon_B^r/f_B^r = 4 \times
{10}^{-3}$, and $A_{*}=0.4$, inferred in Refs. \cite{Kum1}); 
$N_{\mu} \sim {10}^{-3}$ events. 
For SSC, $\mathcal{E}_{\gamma 2}^{\rm{iso}} = {10}^{54.5} \, \rm{ergs}$, 
$\alpha=0.86$, $\beta=3.6$, and $\xi_{\rm{acc}}=1$. 
Photosphere (baryonic) (thick dashed line):
$r=r_{\rm{ph}}={10}^{12.5}$ cm; $N_{\mu} \sim 0.2$ events. 
Photosphere (pair) (thin dashed line):
$r=r_{\rm{ph}} = {10}^{14.2}$ cm; $N_{\mu} \sim 2 \times {10}^{-3}$ events. 
Subphotosphere (baryonic) (thick dotted line):
$r={10}^{-0.5} r_{\rm{ph}}$; $N_{\mu} \sim 0.2$ events.
Subphotosphere (pair) (thin dotted line):
$r={10}^{-0.5} r_{\rm{ph}}$; $N_{\mu} \sim {10}^{-2}$ events. 
For photosphere and subphotosphere, $\mathcal{E}_{\gamma}^{\rm{iso}} = 
{10}^{54.5} \, \rm{ergs}$,  $\xi_B=1$, and $\xi_{\rm{acc}}=1$.
}}
\end{figure}

In Fig. 2, we show neutrino spectra in the SSC and photospheric 
scenarios for an energetic burst such as GRB 080319B. 
In the SSC scenario, expected event rates are few due to the small 
photon density at large radii. Note that neutrinos
coming from protons accelerated at the reverse shock are dominant at
lower energies [see SSC (RS-IS)]. It is because reverse-shock
protons interact with blue-shifted prompt photons, which are
assumed to have the steep photon spectrum with $\beta=3.6$
(solid lines) \cite{Kum1}. 
In the photospheric scenario, we can expect
higher neutrino fluences than in the other scenarios. 

It is important to consider the cumulative neutrino background, since 
the time- and space-coincidence is expected for GRB neutrino
emission \cite{Der1}. 
The background flux can be roughly estimated as \cite{Wax1,KM1,Der1}
\begin{eqnarray}
E_{\nu}^2 \Phi_{\nu} &\sim&  
3 \times 10^{-9} {\rm{GeV cm^{-2} s^{-1} str^{-1}}} \, 
\tilde{\mathcal{E}}_{\rm{HECR},51} \nonumber\\
&\times&  \frac{f_{\rm{mes}}(E_{\nu})}{0.5}
\frac{f_{\rm{sup}}(E_{\nu})}{0.5}    
\frac{f_{z}}{3}
\frac{R_{\rm{GRB}}(0)}{20 \, \rm{Gpc}^{-3} \,
\rm{yr}^{-1}},
\end{eqnarray}
where $f_{\rm{sup}}$ 
is the suppression factor due to cooling of mesons and muons
\cite{Wax1,Rac1,KM2,Mes3} and $f_{z}$ expresses the 
contribution from the high redshift GRBs \cite{Wax1,KM1}.
Here $R_{\rm{GRB}}(0)$ is the overall (not apparent) 
local rate, where GRBs are regarded as jets with 
$\mathcal{E}_{\gamma} \sim {10}^{51}$ ergs \cite{KM1,Mes1}. 

\begin{figure}[t]
\includegraphics[width=\linewidth]{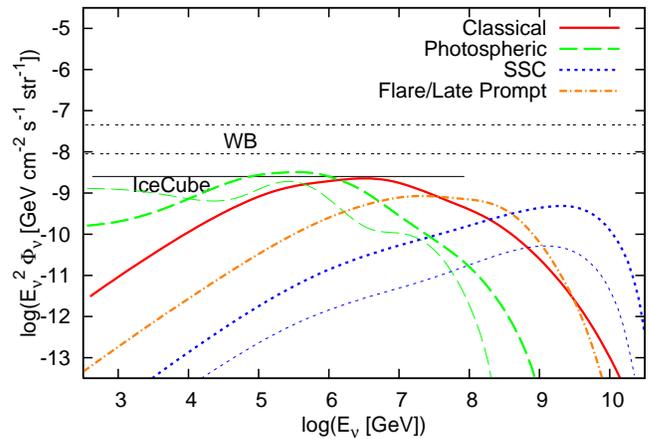}
\caption{\small{\label{Fig3} The cumulative neutrino backgrounds from
GRBs. 
Classical: originally predicted by Waxman and Bahcall \cite{Wax1} and 
the numerically calculated spectrum is taken from Refs. \cite{KM1,KM2}, but 
averaged over sets A and B, $\Gamma={10}^{2.5}$ and $\xi _{B}=1$; 
$N_{\mu} \sim 9.4$ events/yr. 
Photospheric (baryonic) (thick dashed line): 
$r=r_{\rm{ph}}={10}^{12.5} \, \rm{cm}$, $\Gamma={10}^{2.5}$, and $\xi
_{B}=1$; $N_{\mu} \sim 24$ events/yr. 
Photospheric (baryonic) (thin dashed line): 
$r={10}^{-0.5} r_{\rm{ph}}$, $\Gamma={10}^{2.5}$, and $\xi
_{B}=1$; $N_{\mu} \sim 16$ events/yr.
SSC (IS-IS) (thick dotted line):
$r ={10}^{15.5} \, \rm{cm}$, $\Gamma={10}^{2.5}$, 
$L_{\gamma}^b ={10}^{51.3}\, \rm{ergs} \, {s}^{-1}$, and 
$\xi _{B}=0.01$; $N_{\mu} \sim 0.14$ events/yr.
SSC (RS-IS) (thin dotted line):
$r_{\times} \simeq {10}^{16.9}$ cm, $\Gamma_{\times} \simeq 160$, 
and $B_{\times}^{r} \simeq 5.9$ G; $N_{\mu} \sim 0.014$ events/yr. 
Flare/late prompt: proposed by Murase and Nagataki \cite{KM1} and 
the numerically calculated spectrum is taken from Refs. \cite{KM1,KM2} (the model LP0);
$N_{\mu} \sim 1.2$ events/yr. 
WB: Waxman-Bahcall bounds shown as benchmarks \cite{Wax1}.
The cosmic-ray proton energy input per logarithmic interval 
$\tilde{\mathcal{E}}_{\rm{HECR}} \equiv \mathcal{E}_{\rm{CR}}/R$ 
is normalized to ${10}^{51}$ ergs for
the prompt emission scenarios while ${10}^{50}$ ergs for flare/late prompt. 
As a GRB rate evolution model, the GRB3 model in Ref. \cite{KM2} 
is adopted with the cosmological parameters ($\Omega _{\rm{m}}=0.3, 
\Omega _{\Lambda}=0.7; H_{0}=71 \, \rm{km\, s^{-1} \, Mpc^{-1}}$)
and $z_{\rm{max}}=11$. 
}}
\end{figure}

The results are shown in Fig. 3, where the GRB
rate evolution is properly considered \cite{KM2}.
Predictions in the photospheric and SSC scenarios are rather different 
from the classical one.  
In the photospheric scenario, we expect larger event rates than
others with the same baryon loading, and $pp$ neutrinos may become
important at $\lesssim (10-100)$ TeV. In the SSC scenario, 
we could expect $\sim (1-10)$ EeV neutrinos from interactions between
UHECRs and soft photons, but note that our
evaluation would be applied to only a fraction of GRBs \cite{Rom1}.

\section{\label{sec:level5}Implications and Discussions}
In the near future, high-energy neutrino signals may be detected by 
IceCube and KM3Net \cite{Ahr1}. IceCube Deep Core for $\sim$ TeV
neutrinos and acoustic/radio and shower detectors for $\sim$ EeV
neutrinos will also be useful. 
In this work, we have first demonstrated that $pp$ neutrinos can become
more important than $p \gamma$ neutrinos in prompt neutrino emission, 
if the photospheric scenario is realized.
Detection of $\lesssim (10-100)$ TeV neutrinos is important as a probe of 
dissipation and baryon acceleration around/below the
photosphere as well as a diagnosis of the jet composition, 
although the detectability depends on the pair loading. Also, these photospheric
neutrinos can be distinguished from precursor ones \cite{Mes3}
through correlations of neutrinos with prompt photons. 
At higher energies, $p \gamma$ neutrinos become more important, and 
there may be contributions from optically thin internal shocks
occurring above the photosphere. However, 
$p \gamma$ neutrinos cannot be expected for too large values of
$\eta$, although we have assumed $\eta \sim 10$ so far. When
$\eta \gtrsim {10}^{3-4}$, only $pp$ neutrinos may be relevant. 
Note that our predicted fluxes are below the current AMANDA limit 
and Waxman-Bahcall bounds \cite{Ach1}. If neutrinos are not detected in the 
future, it implies that baryon acceleration is insufficient 
(small $\xi_{\rm acc}$ and/or very large $\eta$) or that 
prompt emission occurs sufficiently above the photosphere.
In addition, we have demonstrated that synchrotron photons in the 
optical/ultraviolet range may enhance $\sim$ EeV neutrinos in the SSC
scenario, compared to the other scenarios. Cooling of meson and
muons is not so important, and $\sim$ EeV 
neutrinos will be useful as a probe of UHECR acceleration.
Detection of them will also imply the nature of GRBs, 
e.g., $r$ and $\xi_B$.

One may expect hadronic high-energy gamma rays can be detected by the recently launched 
\textit{Fermi} satellite. For photospheric emission, we do not expect that high-energy 
gamma rays with $\varepsilon_{\rm ob} \gtrsim$ a few $\times \Gamma m_e c^2$ escape 
from the source due to the large optical depth for 
pair creation \cite{Lit1}. Primary hadronic gamma rays induce the cascade 
in the source, and resulting spectra would be similar to those in the classical scenario. 
However, as indicated in Asano and Inoue \cite{Der2}, it is not easy to find 
observational signatures of hadronic gamma rays 
with $\xi_B \sim 1$, $\xi_{\rm acc} \sim 1$, $\alpha=1$ and $\beta=2.2$, considered by us. 
(Note that the steeper electron spectral index of $p_e \sim 3.0$ 
and softer low-energy photon index of $\alpha \sim 1.5$ in their calculations 
could lead to overestimating hadronic signatures in spectra \cite{KM3}.) 
Furthermore, we would not expect to distinguish between the classical and photospheric 
scenarios from signatures of cascaded hadronic gamma rays themselves. Hence, detection of 
$\lesssim (10-100)$ TeV $pp$ neutrinos is more important as a unique probe. 
On the other hand, in the SSC scenario, primary high-energy gamma rays produced via the 
photomeson production could escape from the source. Their typical 
energy is $\gtrsim$ EeV, and they cannot avoid attenuation by the 
cosmic background photons. The detectability of cascaded
gamma rays in the GeV-TeV range strongly depends on the intergalactic magnetic field strength.   
Only when the intergalactic magnetic field is very weak, $B_{\rm IG} \lesssim 
{10}^{-16}$ G, we may detect them as a pair echo \cite{Raz1}. 
However, even in such cases, we expect that the pair echo emission would 
be subdominant compared to the primary leptonic emission. 
In fact, in the SSC scenario, strong high-energy gamma rays should be 
generated through the SSC emission, and the isotropic radiation energy 
around the third peak, ${\mathcal{E}}_{\gamma 3}^{\rm iso} \sim 
{10}^{54.5}~{\rm ergs}$, is larger than the isotropic energy of 
primary hadronic very high-energy gamma rays, 
${\mathcal{E}}_{\gamma, \rm VHE}^{\rm iso} \sim {10}^{53.5}~{\rm ergs}$. 
Hence, it would be difficult to use hadronic gamma rays as a probe of 
baryon acceleration in the SSC scenario.

GRBs may be the origin of observed UHECRs, as is discussed 
in many papers. Note that, if the photospheric scenario is real, 
UHECR explanation is impossible due to strong proton losses at the
highest energies. In the SSC scenario, UHECRs may be explained.

\begin{acknowledgements}
As this work was being completed, we became aware
during the Nanjing GRB conference 
that Wang and Dai also studied subphotospheric emission 
independently, albeit with a simpler analytic formulation \cite{Wan1}.  
K.~M. thanks X.~Y. Wang, P. M\'esz\'aros, K. Asano and Y. Z. Fan.
K.~M. is supported by a Grant-in-Aid for JSPS.
\end{acknowledgements}


\newpage 

\end{document}